\documentclass[sigconf]{acmart}
\AtBeginDocument{%
  }

\copyrightyear{2025}
\acmYear{2025}
\setcopyright{rightsretained}
\acmConference[CHI EA '25]{Extended Abstracts of the CHI Conference on Human Factors in Computing Systems}{April 26-May 1, 2025}{Yokohama, Japan}
\acmBooktitle{Extended Abstracts of the CHI Conference on Human Factors in Computing Systems (CHI EA '25), April 26-May 1, 2025, Yokohama, Japan}
\acmDOI{10.1145/3706599.3720227}
\acmISBN{979-8-4007-1395-8/2025/04}

\usepackage{graphicx}
\usepackage{color}
\usepackage{xcolor}
\usepackage{pifont}
\usepackage{xspace}

\newcommand{\custpara}[1]{\vspace{.1cm}\noindent\textbf{#1}\space}
\newcommand{\RLjam}{ReaLJam\xspace}

\begin{document}

\title[ReaLJam: Real-Time Human-AI Music Jamming with RL-Tuned Transformers]{ReaLJam: Real-Time Human-AI Music Jamming with Reinforcement Learning-Tuned Transformers}

\author{Alexander Scarlatos}
\email{ajscarlatos@cs.umass.edu}
\orcid{0000-0001-6419-5909}
\authornote{Work done during internships at Google DeepMind.}
\affiliation{%
  \institution{University of Massachusetts Amherst}
  \country{United States}
}

\author{Yusong Wu}
\email{yusong.wu@umontreal.ca}
\authornotemark[1]
\affiliation{%
  \institution{University of Montreal, Mila}
  \country{United States}
 }

\author{Ian Simon}
\email{iansimon@google.com}
\affiliation{%
  \institution{Google DeepMind}
  \country{United States}
}

\author{Adam Roberts}
\email{adarob@google.com}
\affiliation{%
 \institution{Google DeepMind}
 \country{United States}
}

\author{Tim Cooijmans}
\email{cooijmans.tim@gmail.com}
\affiliation{%
  \institution{University of Montreal, Mila}
  \country{United States}
}

\author{Natasha Jaques}
\email{natashajaques@google.com}
\affiliation{%
  \institution{Google DeepMind}
  \institution{University of Washington}
  \country{United States}
}

\author{Cassie Tarakajian}
\email{ctarakajian@google.com}
\affiliation{%
 \institution{Google DeepMind}
 \country{United States}
}

\author{Cheng-Zhi Anna Huang}
\email{annahuang@google.com}
\affiliation{%
 \institution{Google DeepMind}
 \country{United States}
}

\renewcommand{\shortauthors}{Scarlatos et al.}

\begin{abstract}
Recent advances in generative artificial intelligence (AI) have created models capable of high-quality musical content generation. However, little consideration is given to how to use these models for real-time or cooperative jamming musical applications because of crucial required features: low latency, the ability to communicate planned actions, and the ability to adapt to user input in real-time. To support these needs, we introduce \RLjam, an interface and protocol for live musical jamming sessions between a human and a Transformer-based AI agent trained with reinforcement learning. We enable real-time interactions using the concept of anticipation, where the agent continually predicts how the performance will unfold and visually conveys its plan to the user. We conduct a user study where experienced musicians jam in real-time with the agent through \RLjam. Our results demonstrate that \RLjam enables enjoyable and musically interesting sessions, and we uncover important takeaways for future work.
\end{abstract}


\begin{teaserfigure}
  \centering
  \includegraphics[width=.6\textwidth]{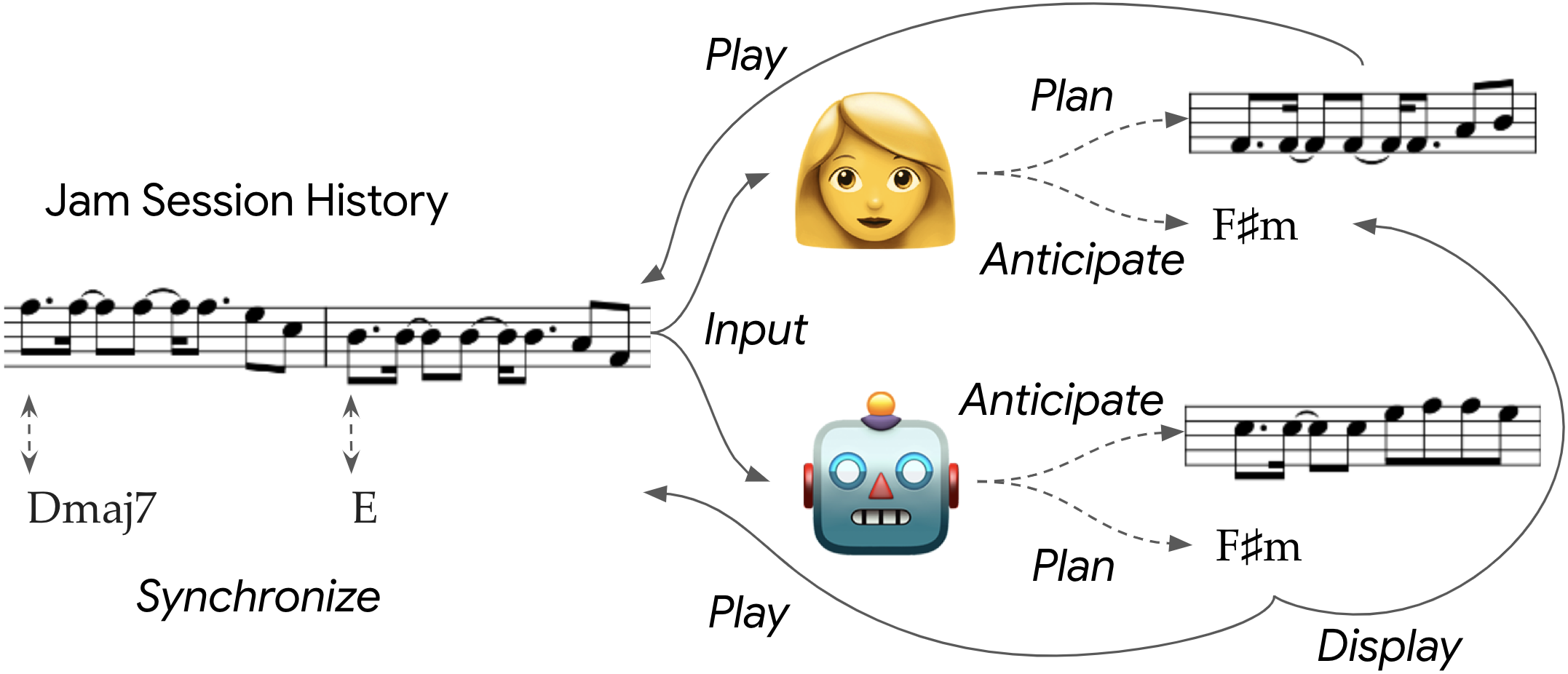}
  \caption{The live jamming task. Based on the session history, the agent anticipates the user's melody and plans what chords to play. Similarly, the user plans what melody to play, but anticipates chords by viewing the agent's plan displayed onscreen. The user and agent play their respective parts, synchronized by the client, to create a live jam.}
  \Description{The left side contains melody notes with chords symbols below, labeled "Jam Session History" with the word "Synchronize" below. The right contains two characters, a human and an AI, which both receive the session history as input. The human plans a melody and anticipates a chord, while the AI anticipates a melody and plans a chord. The AI also displays its planned chord to the human.}
  \label{fig:coadaptation}
\end{teaserfigure}


\maketitle

\section{Introduction}\label{sec:introduction}

While there have been many recent advances in AI-generated music, how to leverage these advances for \textit{live} generation, particularly \textit{cooperatively} with humans, remains understudied. Cooperative human-AI music creation in a live setting has many potential applications: it can assist in the spontaneous creation of ideas, enhance improvised performances, and provide practice partners or musical tutors to people who may not have access to them otherwise. Many of these applications can manifest as a live ``jam'', where a human and AI agent play respective musical parts live, building off each others' musical contributions to form a cohesive improvised performance.

However, utilizing AI for live jamming comes with many challenges: First, in order for the agent to play in a way that is cohesive with the human, the agent must \textbf{anticipate} what the human will play in the short-term future, planning its own performance based on this prediction. This act of anticipation happens naturally for humans \cite{huron2008sweet}, but has also been leveraged in non-live generative music models \cite{thickstun2024anticipatory} to improve the quality of accompaniment. Second, the human must similarly anticipate the agent's future actions. When reading music or playing music games like Guitar Hero \cite{guitar-hero}, humans look ahead to prepare for what to play next. During live performances, humans will signal future plans to each other through intricate gestures via their head, body, and hands \cite{king2016gestures}. Therefore, an effective jamming agent should have a similar way of conveying its plan to the human to help them anticipate upcoming actions. Finally, the interface must \textbf{synchronize} the user and agent notes along the time dimension so they are played at their intended times with no discernible delay. Prior works either accept a delay in system responses (in non-jamming settings) \cite{musicfx} or use small models, such as LSTMs, with unnoticeable delay \cite{bachduet}. However, more powerful Transformer-based music models \cite{music-transformer} have slower inference and run on specialized compute hardware, often requiring them to run remotely on a server and adding further delay to responses.

While there are many prior works on real-time human-AI collaboration in music \cite{mccormack2020design, kivanc2019musical, biles1994genjam, biles1998interactive, pachet2013reflexive, thelle2021spire, musicfx, cococo, expressive-comm, bachduet}, only one studies a real-time jamming system \cite{bachduet}, but does not fully address the above challenges (see a detailed related work in Appendix~\ref{sec:related-work}).
We note that anticipation and synchronization  are relevant in other real-time tasks, such as collaborative drawing \cite{10.1145/3491102.3501914} or speech-to-speech conversation \cite{veluri-etal-2024-beyond}, necessitating solutions for these increasingly important challenges.

\subsection{Contributions}

\custpara{System}
In this work, we introduce \RLjam (Real-Time Reinforcement Learning Jamming), a holistic system that enables live human-AI jamming via a web interface and techniques for interacting with AI music agents in real-time. We specifically study chord accompaniment, where the user plays a melody and a Transformer-based agent plays along with accompanying chords. To the best of our knowledge, our system is the first to achieve live jamming with large Transformer models. We conduct a user study where experienced musicians play multiple performances with \RLjam, validating the system's effectiveness and enjoyment for users while uncovering meaningful findings regarding music modeling and the importance of giving users fine-grained control over interface settings. See video and audio examples of users playing live with \RLjam on our demo page: \url{https://storage.googleapis.com/genjam/index.html}.

\custpara{Anticipation and Synchronization}
Our technical contributions address the challenges of \textbf{anticipation}, i.e., both the user and agent being able to reasonably predict how the other will act, and \textbf{synchronization}, i.e.,  ensuring that user and agent notes are played at the intended times without delay. We propose multiple technical and design solutions to address these challenges, such as using the agent to construct a near-term musical plan and displaying upcoming chords to the user via a waterfall display. Figure~\ref{fig:coadaptation} illustrates the collaborative task between the user and agent, including the roles of anticipation and synchronization.

\section{ReaLJam}

\begin{figure*}
 \centerline{
 \includegraphics[width=\linewidth]{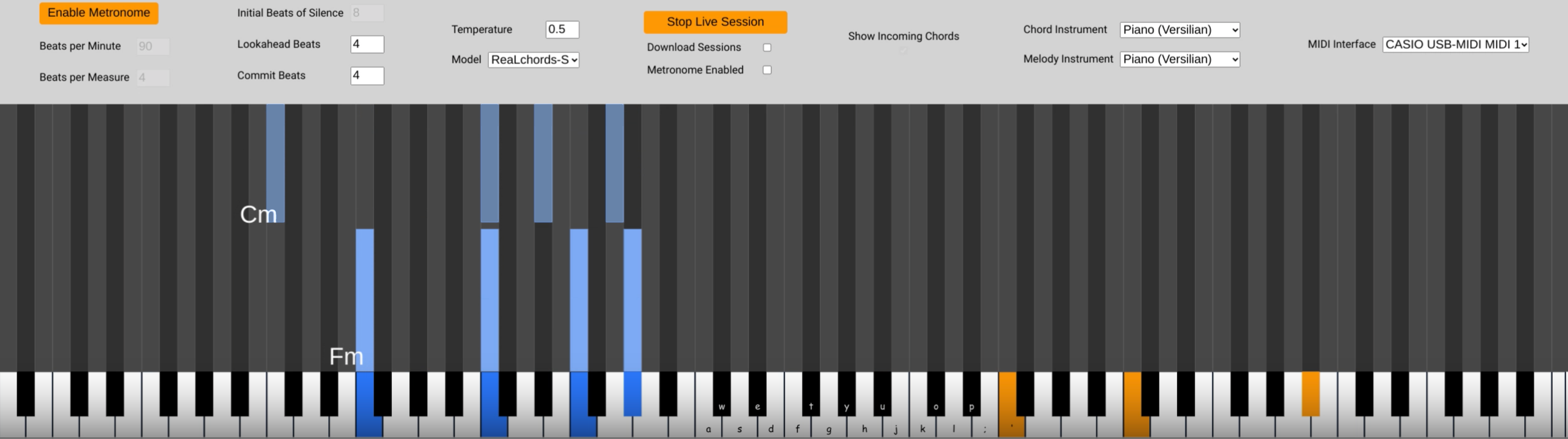}}
 \caption{The \RLjam interface. Users see a control bar at the top and a piano at the bottom. Keys for sustained notes are highlighted for the human (orange) and agent (blue). Anticipated chords fall from the control bar to the piano, with uncommitted chords rendered semi-transparently. See a video of a user playing live with \RLjam on our demo page: \url{https://storage.googleapis.com/genjam/index.html}}
 \Description{Top: a control bar with inputs for "Enable Metronome", "Beats per Minute", "Beats per Measure", "Initial Beats of Silence", "Lookahead Beats", "Commit Beats", "Temperature", "Model", "Stop/Start Live Session", "Download Sessions", "Metronome Enabled", "Show Incoming Chords", "Chord Instrument", "Melody Instrument", and "MIDI Interface". Middle: a grid of dark vertical lines, where in the top half the corresponding notes (in a semi-transparent blue) and the symbol for a C minor chord are shown, and in the bottom half the corresponding notes (in an opaque blue) and the symbol for a F minor chord are shown. Bottom: a piano keyboard, where the notes for a F minor chord are highlighted in blue, and F, C and B flat notes are highlighted higher up in orange.}
 \label{fig:interface}
\end{figure*}

In this section, we detail the various components of \RLjam and how they come together to enable live human-AI jamming.
We address the underlying challenges of \textbf{anticipation} to improve user and agent predictions, and \textbf{synchronization} of the user and agent parts to enable a fluid jamming experience. We detail our solutions to these challenges through the design of the user interface (Section~\ref{sec:user-interface}), the use of an agent trained for online chord generation with reinforcement learning (Section~\ref{sec:model}), and the communication protocol for client-server synchronization (Section~\ref{sec:protocol}). We show a screenshot of a user playing with \RLjam in Figure~\ref{fig:interface}.

\subsection{User Interface}\label{sec:user-interface}

In the \RLjam interface, a user is presented with a piano at the bottom of the screen, a control bar at the top, and a grid region in the middle where upcoming chords will appear. Users play notes with a MIDI device or computer keyboard and see them light up on the piano. This design is similar to the popular piano learning app Synthesia \cite{synthesia}.

\custpara{Agent Interaction}
After the user clicks the button to start a live session, their first note initiates the \textit{silence period}, where an agent will first wait and ``listen'' to user's notes. This period allows the agent to collect sufficient information to make informed predictions by the time it starts generating. After this period, the agent will begin to produce chords. Inspired by waterfall displays in music games like Synthesia \cite{synthesia} and Guitar Hero \cite{guitar-hero}, the user will see the upcoming chords falling down the grid region. When the chords reach the piano, they are played and the associated keys light up in blue. The number of beats into the future which the user can see upcoming chords is called the \textit{lookahead}. Chords in the immediate future up to a \textit{commit time} (before the end of the lookahead) are said to be \textit{committed}; that is, the agent can no longer change those predictions. Beyond the commit time, the agent is free to update its predictions. Uncommitted chords are rendered semi-transparently, and may change until they are committed. The commitment ensures that the user is given enough time to react to the incoming chords, and the user can adjust the commit time for a tradeoff between knowing chords ahead of time and having the chords be more up to date with the melody.

\custpara{User Control}
Following prior work showing the importance of user control in music generation interfaces \cite{cococo}, we provide fine-grained settings for various aspects of our system. Specifically, users can configure the number of beats for the silence period, lookahead, and commit time, whether to show incoming chords, the agent model and sampling temperature, a metronome with tempo and time signature, and digital instruments for the melody and chord parts.

\subsection{Music Generation Agent}\label{sec:model}

\RLjam's chord generation agent is based on ReaLchords \cite{realchords}, recent work which was the first to enable real-time online chord accompaniment generation with a Transformer model by fine-tuning with reinforcement learning (RL). The agent conditions on an input melody and previously generated chords to predict what chord should be played next. Inputs and outputs are discretized into time frames of 1/16th notes, with tokens representing melody notes or chords. ReaLchords includes online and offline models, where the online model is used for real-time generation, seeing only what has been played so far, while the offline model sees the past and future of the melody. Both models are initially pre-trained with a dataset from Hooktheory~\cite{hooktheory} containing $\sim$30,000 annotated pop song snippets, after which the online model undergoes further fine-tuning through RL. This RL training involves generating chords for fixed melodies and constructing a reward based on the offline model and other reward models. We experiment with two RL variants of ReaLchords: \textbf{RL-S}, which evaluates rewards at sequence endpoints, and \textbf{RL-M}, which evaluates rewards throughout the sequence to ensure local coherence, with the pre-trained online model named \textbf{Online}.

\subsection{Communication Protocol}\label{sec:protocol}

\begin{figure*}
    \centering
    \includegraphics[width=.7\linewidth]{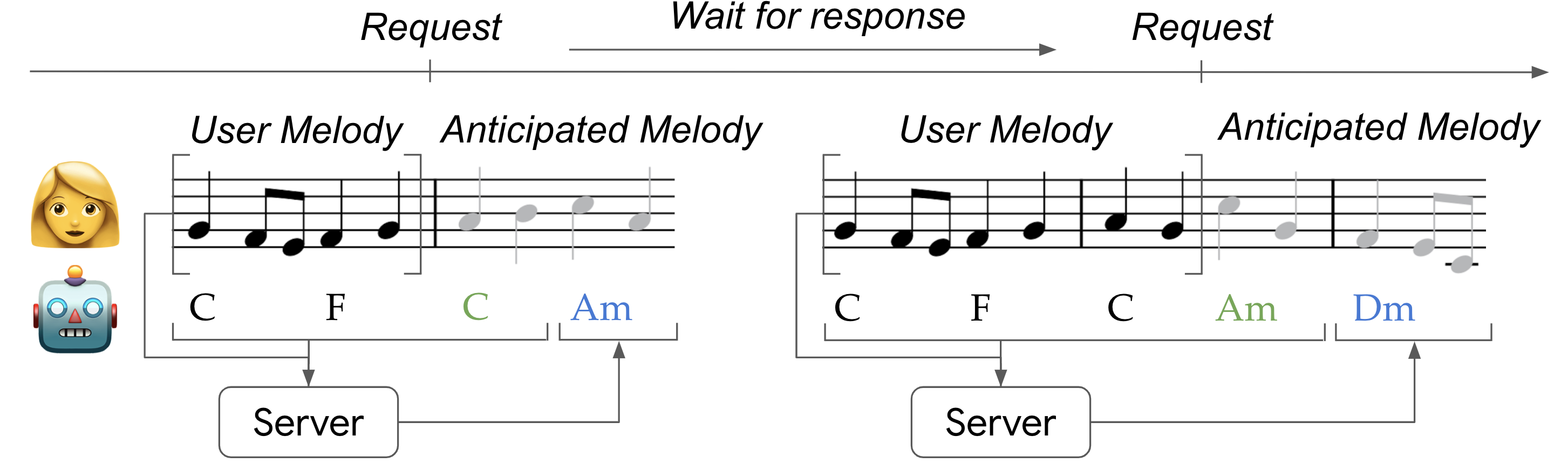}
    \caption{Synchronization in \RLjam is robust to server latency. We show a request/response loop with a 2 beat round-trip latency for demonstration, with a 4 beat lookahead and 2 beat commit time. The client plays cached future chords (green and blue) while waiting for responses. Each request, the client sends the server the played melody (black), played chords (black), and committed chords (green). The agent predicts chords several frames into the future by anticipating user input (gray) with committed chords unchanged. Upon receiving the response, the client schedules new chords (blue) after the commit period and sends a new request.}
    \Description{Top: a timeline reads "Request", "Wait for response" (with an arrow underneath) leading to another "Request". Bottom: two panels show two time frames, one corresponding to each request/response. At the top of each panel is a music bar showing dark notes labeled "User Melody" on the left and gray notes labeled "Anticipated Melody" on the right. Below the music bar are corresponding chords in black, followed by green and light blue. Below each panel is a "Server" box that takes as input the dark notes/chords and green chords, and outputs the blue chords and gray notes.}
    \label{fig:timeline}
\end{figure*}

In order for users to perceive that the agent is immediately reacting to their input, we introduce a protocol for real-time synchronization between the web client running the interface (Section \ref{sec:client}) and a server-hosted agent (Section \ref{sec:server}).
We illustrate an example synchronization timeline in Figure~\ref{fig:timeline}.

\subsubsection{Client}
\label{sec:client}
\RLjam's interface runs in a web client, where continuous requests are made to a stateless server to get up-to-date chord predictions for the near-term future. The client performs the following in a persistent loop:

\custpara{Request}
Each request, the client sends the entire jam session history (melody and chords) plus chords currently in the lookahead, the target frame to start generating in, and all relevant settings such as the number of frames for lookahead, commit, etc. The chords in the request are simply tokens copied from the previous response.

\custpara{Response}
After receiving the server's response, the client schedules the returned chords in the lookahead. All chords previously scheduled to play after the current frame are canceled and replaced by the chords in the response. The client uses the chords' associated frames in the response to schedule them at the correct times based on the beats per minute and the session start time, achieving time synchronization. In cases where the response takes longer than one frame, scheduled chords play while waiting, ensuring no client-side delay or interruption as long as the number of lookahead frames is higher than the number of frames spent waiting. In practice, using a single device for inference, a majority of responses return within 100 milliseconds, fast enough for single-frame round trips at 150 beats per minute. After the response is processed, the client makes a new request for the next frame, continuing until the user ends the session.

\subsubsection{Server}
\label{sec:server}
The server's job is to run the agent to generate chords for the lookahead given the current session history. For each request it receives, the server performs the following steps in order:

\custpara{Tokenize Melody}
The server converts the melody in the request to a monophony so that at most one note is registered in each frame. Within each frame, we simply take the first note played and ignore any others.

\custpara{Warm-Start Generation}
To improve chord quality early in the session, we warm-start online generation with offline-generated chords in the silence period. Once per session, 4 frames before the \textit{silence period} ends, we use the \textit{offline} model to generate chords to accompany the melody so far. We do not use the online model for this step since it cannot see the full melody in the silence period, whereas the offline model can. The offline-generated chords are returned to the client to be echoed back for future generations, but are not played by the client. This design is motivated by previous work that conditions online models with ground-truth data \cite{realchords, rl-duet}.

\custpara{Commit Period}
For the frames in the commit period, we fix the previously predicted chords (sent in the request) but anticipate notes in the user melody that will play over these chords. 
We do this by having ReaLchords generate chord and melody tokens up to the last committed frame of the lookahead (identified in the request), conditioned on the last 512 tokens of the jam session history. We then replace these generated chords with the previously committed chords.

\custpara{Adaptive Period}
Finally, we generate chords for the adaptive period, i.e., the frames between the end of the commit period and the end of the lookahead. Conditioned on the jam session history, the committed chords, and the melody anticipated in the previous step, we use ReaLchords to generate chord and melody tokens up to the last frame of the lookahead. By doing so, the agent concurrently anticipates user actions while constructing a plan for its own actions. All chords in this new lookahead are then converted from symbolic form to pitches and returned to the client.

\section{User Study}

To investigate \RLjam's viability for live human-AI jamming and the quality of its user experience, we conducted a user study where 6 participants, each an experienced musician, play multiple live performances with \RLjam. While this is a small participant pool, we conduct extensive interviews with each, collecting a high amount of quality data for in-depth analysis. We find that our study is sufficient to validate the effectiveness of our system, identify its strengths and weaknesses, and discover important directions for future work. We now detail our study design.

\custpara{Participants}
For recruitment, we posted calls for volunteer participation in music interest groups within our institution, where the only prerequisites for participants were to have some piano experience and be comfortable improvising. Candidates applied via email and were selected on a first-come-first-serve basis. We conducted a trial version of the study with 2 participants, after which we refined the study and conducted a final version with 6 participants.

\custpara{Study Overview}
For each participant, we carry out a 1 hour session over a video call, during which participants perform with \RLjam and answer questions.
After we explain the interface, participants practice with \RLjam to get a feel for the interaction and set the tempo to their preference. When participants are ready, they play a 45-second \textbf{baseline} performance with \RLjam. Afterwards, they play 8 45-second performances, where in each one a single interface setting differs from the baseline. Finally, participants play one last 45-second performance using any settings they like. After the video call, participants answer a survey to provide aggregated quantitative and qualitative data.

\custpara{Setting Experiments}
We examine the impact of various settings for the \textbf{User Interface}: whether incoming chords are shown (\textbf{I.C.}) and whether the metronome is on (\textbf{Met.}), settings for the \textbf{Agent}: the underlying \textbf{Model} used (between RL-S, RL-M, and Online) and the sampling temperature (\textbf{Temp.}) (between 1 and 0), and settings for \textbf{Synchronization}: the number of beats to commit (\textbf{Com.}) (between 4, 2, and 0) and the initial beats of silence (\textbf{I.S.}) (between 8 and 0). The lookahead is always 4 beats. We always use a commit time of 0 for the Online model since it is highly sporadic otherwise, likely due to a lack of ability to plan ahead induced by RL training. The settings for the baseline performance are as follows: incoming chords are shown, the metronome is on, the agent is RL-S, the temperature is 1, the commit is 4 beats, and the initial silence is 8 beats. For each subsequent performance, we modify exactly one setting from the baseline and fix the rest. We randomize the order of experiments for each participant to avoid bias, with the constraints that commit must be tested before models
and the two commit tests will occur consecutively as will the two model tests.

\custpara{Intra-Study Metrics}
For each performance following the baseline, we ask participants 4 questions regarding how their experience changed relative to the baseline: ``Which performance was more musically interesting?''\ (\textbf{Int.}), ``In which performance did the model adapt to your melody better?''\ (\textbf{Adpt.}), ``In which performance did you feel more in control?''\ (\textbf{Ctrl.}), and ``Which performance did you enjoy playing more?''\ (\textbf{Enj.}). For each question, participants identify which performance was better or can respond with ``No Difference.'' We chose to ask for comparisons rather than absolute ratings since the former have higher validity and are more accurate (e.g. \cite{goffin2011all,vidulich1987absolute,jones2015peer}).

\custpara{Post-Study Metrics}
After their sessions, we sent participants a survey with questions on their musical background and impressions of their experience with \RLjam.
We asked the users the following quantitative questions, all using a 1-5 Likert scale: For user background, we asked ``How experienced are you at the piano?''\ (\textbf{Exp. Piano}) and ``How experienced are you at improvising music with other people?''\ (\textbf{Exp. Improv.}). Having users reflect on when the system was working at its best, we asked them ``How musically interesting/exciting (in terms of harmonization, rhythm, etc.) was the session?''\ (\textbf{Int.}), ``How well did the chord model adapt to your melody?''\ (\textbf{Adpt.}), ``How much control/agency did you feel during the session?''\ (\textbf{Ctrl.}), and ``How much did you enjoy the jamming experience?''\ (\textbf{Enj.}). Regarding their experience overall, we asked users ``How often would you use this software outside of the experimental setting?''\ (\textbf{Util.}).
We also asked users the following open-ended questions regarding their experience overall: ``What were some things that you enjoyed most about the jamming experience?'',\ ``What were some things that you disliked about the experience?'',\ ``In what ways was jamming with the AI similar to or different than jamming with a real human?'',\ ``Do you have any suggestions for improving the jamming experience?'',\ and ``Do you have any other thoughts or comments?''.

\section{Results and Discussion}

\begin{table*}[]
    \centering
    \small
    \begin{tabular}{cc|ccccc}
        \multicolumn{7}{c}{\textbf{Overall Scores}}\\
        \textbf{Exp. Piano} & \textbf{Exp. Improv.} & \textbf{Int.} & \textbf{Adpt.} & \textbf{Ctrl.} & \textbf{Enj.} & \textbf{Util.}\\
        \hline
        3.3 & 3.8 & 3.5 & 2.7 & 2.7 & 4.3 & 3.2\\
    \end{tabular}
    \quad
    \begin{tabular}{lcccccccc}
        \multicolumn{9}{c}{\textbf{Average System Setting Comparison Scores}}\\
        & \textbf{I.C. On} & \textbf{Met. On} & \textbf{RL-M} & \textbf{Online} & \textbf{Temp. 1} & \textbf{Com. 2} & \textbf{Com. 0} & \textbf{I.S. 8} \\
        \hline
        Avg. Preference & -0.21 & 0.17 & 0.29 & -0.58 & -0.08 & 0.08 & 0.21 & 0.25\\
        Chosen in Last & 5/6 & 3/6 & 3/6 & 0/6 & 5/6 & 1/6 & 3/6 & 5/6
    \end{tabular}
    \caption{\textbf{Top:} Users' prior experience and average overall scores for the system, each in the range $[1,5]$. Users have high levels of prior experience and enjoyed using \RLjam. \textbf{Bottom:} Users' intra-study experiment scores for each setting, averaged over users and all 4 preference questions. A user choosing the indicated setting gives a score of 1, choosing ``No Difference'' gives a score of 0, and choosing the alternative gives a score of -1. The Online model is greatly dispreferred to the RL models, whereas other settings are mostly balanced across users and measures. We also show the portion of times each setting was chosen by users for their last performance, where most users chose to see incoming chords and include the silence period, two of our key technical contributions.}
    \label{tab:quant_results}
\end{table*}

We now detail our main findings by analyzing data from the intra-study questions and the post-study survey, behaviors from participants during the sessions, and musical artifacts collected from the performances. We find that 1) \RLjam successfully enables users to jam with an AI in an enjoyable and meaningful way, 2) RL agents greatly improve musical quality but lack some desirable behaviors, and 3) user experience is highly impacted by interface settings with regard to personal preference. We summarize the quantitative results in Table~\ref{tab:quant_results} with a more detailed breakdown Appendix~\ref{sec:intra-metrics}. Our findings are further supported by audio examples in our demo page: \url{https://storage.googleapis.com/genjam/index.html}.

\subsection{\RLjam successfully enables human-AI jamming}

\custpara{A real-time system that works} Users were very positive about their experiences live jamming with an AI for the first time. One user said ``It was nice to be able to jam without scheduling with another person'' (P4). Others noted the ``low latency'' (P1), ``ease of setup/integration'' (P1), and ``low overhead to just jamming away'' (P6). This sentiment shows that our client-server synchronization successfully supports a seamless and low-latency jamming experience.

\custpara{Exceptional musical moments} Another positive factor for users was having moments when the agent surprised them with especially cohesive or interesting chords. When asked what they liked about the experience, users noted ``surprise! Moments of unexpected joy'' (P1), ``It kept me on my toes [with respect to] which chord comes next'' (P3), and ``There were several times where the model did predict the chord I had in my mind - those moments were really the strong positive moments in the whole experience'' (P5). These moments contributed to the Interesting and Enjoyable scores of 3.5 and 4.3, respectively. These results are especially encouraging considering that users reported high levels of experience in piano and improvisation, indicating that the user experience is approved by a knowledgeable audience.

\custpara{Future use} Users also expressed that they would like to use a tool similar to \RLjam in their workflows. One user said, ``This was fun, would be great to have it as a MIDI plugin or DAW tool'' (P2), and another, ``I could see this being more useful as a compositional tool in `here’s a melody, and I have chords, but I’m not going to tell you and want to hear your chords you think fit the entire melody'. That would be pretty useful in my workflow'' (P6). Users also gave an average Utility score of 3.2, indicating  \RLjam could be useful to musicians in real-world settings.

\subsection{Reinforcement learning is necessary}

\custpara{RL agents significantly outperform pre-trained models} When comparing the ReaLchords-S (RL) and Online (pre-trained) models, users preferred ReaLchords-S in all 4 metrics by a large margin. This is mainly because the Online model frequently entered failure modes; it would generate mostly coherent chords for some time, but then begin generating sporadic, unpredictable chords and extended periods of silence. This can be explained by the Online model not handling out-of-distribution data well and not being able to recover from random variations in user input or its own generations, similar to findings in prior work \cite{realchords}. We also observe that ReaLchords-M is generally preferred by users, outperforming ReaLchords-S on all 4 categories by a small margin and being chosen by half of users in their last session. This indicates that not only is RL important, but the choice of reward function can have a meaningful impact on users.

\custpara{Best agents still have shortcomings} While the RL agents generally enabled an enjoyable musical experience, users expressed two main issues with the chord predictions: they did not always match what users were expecting, and the agent had little sense of musical structure. Users noted that chords sometimes felt ``random'' (P1, P2), and ``It was more fun playing with a model that seemed to `get' what I was trying to do, and less fun when I felt like it was just guessing'' (P5). Additionally, users noted ``little (if any) regard to song structure'' (P3), the agent struggling to ``pick up on the `4 bar pattern''' (P4), and there ``didn't seem to be any inherent desire to resolve on bar 8'' (P6). These issues could likely be addressed with new reward functions, such as structure-based \cite{rnn-rl, sancaktar2023regularity} or diversity-based \cite{diverse-reward} rewards.

\custpara{Techniques to improve chord quality}
By observing musical artifacts created during the user study, we find that the silence period and longer commit periods can improve chord quality. Without the silence period, the initial first chord is almost always dissonant with the melody, whereas allowing the agent to listen for 8 beats almost always makes the first chord harmonic with the melody. Additionally, when the number of commit beats is 0, there are often rapid unnatural chord changes due to the agent changing its plan. Using 4 beats of commit greatly improves the plan's stability.

\subsection{Settings greatly impact user experience}

\custpara{Preferred settings are highly user specific} Perhaps surprisingly, we found that users cared greatly about the particular interface settings yet also disagreed with each other about which settings were best. These effects were strongest with showing incoming chords, the commit beats, and the metronome. For incoming chords, P1 indicated they enjoyed seeing incoming chords because it reminded them of getting visual feedback from humans and P6 struggled to synchronize their melody with the chords when they weren't shown, yet P5 generally didn't even look at the screen when the chords were shown. This discrepancy indicates that while assisting with anticipation can be helpful for some users, others may be able to sufficiently predict agent actions on their own. For commit, some users felt setting the value to 0 allowed for more accurate predictions by being able to generate chords ``just in time'' (P2), while another felt that setting it to 0 made it ``step on your toes'' too much by changing chords unpredictably (P1). For the metronome, some users were unable to time their melodies well without it (P6), while another felt it made them feel too ``boxed in'' (P5).
Overall, these results make it clear that it is critical to give users a high level of control over interface settings, due to their diverse personal preferences and musical styles, consistent with findings in prior work \cite{cococo}.

\custpara{User preference transcends objective measures}
We find that the settings users preferred did not always match the objective ratings given, indicating that user preference is complex and further motivating high levels of user control in musical interfaces. For instance, users generally agreed that not seeing the incoming chords made the session more musically interesting, likely because of the surprisal factor of not knowing what would come next. However, 5/6 users still chose to see incoming chords in the last performance, indicating a tradeoff between musical interest and ease of anticipating the agent's chords. Similarly, users thought that a temperature of 0 made the agent more adaptive due to less random predictions, but 5/6 ultimately set temperature to 1 to increase chord diversity.

\section{Conclusion}

In this paper, we introduce \RLjam, a system for real-time musical jamming between users and AI agents. We propose utilizing anticipation to improve predictions for both the user and agent, which involves making a future plan with the agent and conveying the plan to the user via a waterfall display. We also propose a communication protocol to synchronize the user and agent musical parts in real-time by further utilizing the agent's future plan. We conduct a user study where musicians jam with \RLjam and reflect on their experiences, and find that \RLjam successfully enables live jamming with a high level of enjoyment, musical interest and utility.

There are many avenues for future work. First, our findings from the user study can be used to inform future generative online music model designs. In particular, rewards should be developed to encourage high-level musical structure and adapt to a diversity of melodic styles. Second, more high-level control over the agent should be given to users, such as specifying the musical genre or desired sparsity of model outputs. Third, additional need finding user studies should be carried out to determine how \RLjam can best support both expert and novice musicians, informing future design decisions. Finally, our solutions for anticipation and synchronization can be integrated into other applications for real-time human-AI collaboration, both for music and other creative domains.


\begin{acks}

We would like to thank Jaewook Lee and Noah Constant for their feedback on the paper, the anonymous reviewers for their suggestions, and the rest of the Google Magenta team for their advice and support throughout the project.

\end{acks}

\bibliographystyle{ACM-Reference-Format}
\bibliography{references}

\appendix

\section{Ethics Statement}

There are several potential positive societal outcomes related to our work, while there are ethical risks to consider as well. Primarily, providing people with a high quality automated jam partner could increase access to artistic creation and education, since many people do not have access to human jam partners due to finances, inability to travel, or other reasons. On the other hand, it is possible that automated jam partners could replace paid musicians to some extent, which is a concern across AI domains. Also, in our work, we use models trained on a dataset of primarily Western pop music, which may make our system less accessible to people from other cultures. However, we note that our interface is designed to work with any frame-based model that can anticipate future actions, allowing more culturally diverse models to be applied in future work.

\section{Related Work}
\label{sec:related-work}
There is a long history to designing intelligent music systems that adapt to musicians in real-time~\cite{mccormack2020design, kivanc2019musical}, enabling a wide range of human-AI relationships in music. GenJam~\cite{biles1994genjam, biles1998interactive} trades fours and eights with a musician. Reflexive Loopers~\cite{pachet2013reflexive} extends loop pedals by adapting its choice of samples to features of a musician's performance. Spiremuse~\cite{thelle2021spire} serves as a brainstorming partner which retrieves or recombines musical responses that shadow, mirror or couple with the musician. MusicFX DJ~\cite{musicfx} allows users to add and balance text prompts as music is being generated. Cococo~\cite{cococo} fills in musical parts based on user settings on high-level aspects like emotion, and \cite{expressive-comm} provides users with options for subsequent generations based on previous user selections. However, all these real-time interactions are facilitated through turn-taking, built-in delay, or loose coupling, therefore being unsuitable for the jamming setting where users and agents create music concurrently with no discernible delay.
 
In contrast, we focus on the tightly coupled setting of real-time accompaniment, which requires the machine to be highly synchronized with the musician. Prior work in accompaniment generation often assumes that the system is offline~\cite{huang2019bach, donahue2023singsong, simon2008mysong} or that a pre-defined plan is given, such as a Jazz standard~\cite{nika2012improtek, nika2017improtek}. For the real-time setting, we need to additionally account for delay and also planning. This requires not just generative agents that support online generation and have anticipatory capabilities but also user interfaces that can communicate to a user how an agent is planning and re-planning its trajectory. 

RL-Duet~\cite{rl-duet}, SongDriver~\cite{wang2022songdriver}, and  ReaLchords~\cite{realchords} are generative models and agents that target online generation. However, they have only been tested in simulation and not yet with users, do not run in real-time, and do not have a user interface. To the best of our knowledge, only BachDuet~\cite{bachduet} built a complete interactive system for real-time counterpoint accompaniment. Compared to \RLjam, the implementation of BachDuet is simpler and more genre-specific: it is trained on two-part Bach chorales, uses a locally-run LSTM model, and trains with a supervised objective. In comparison, \RLjam uses the ReaLchords model, a Transformer trained on a large corpus of pop music using RL fine-tuning, and interacts with the model through a client-server architecture.

\onecolumn

\section{Detailed Intra-Study Results}
\label{sec:intra-metrics}

\begin{table*}[h]
    \centering
    \small
    \begin{tabular}{l|cccc|cccc|cccc|cccc}
        & \multicolumn{4}{c|}{\textbf{I.C. On}} & \multicolumn{4}{c|}{\textbf{Met. On}} & \multicolumn{4}{c|}{\textbf{RL-M}} & \multicolumn{4}{c}{\textbf{Online}}\\
        & \textbf{Int.} & \textbf{Adpt.} & \textbf{Ctrl.} & \textbf{Enj.} & \textbf{Int.} & \textbf{Adpt.} & \textbf{Ctrl.} & \textbf{Enj.} & \textbf{Int.} & \textbf{Adpt.} & \textbf{Ctrl.} & \textbf{Enj.} & \textbf{Int.} & \textbf{Adpt.} & \textbf{Ctrl.} & \textbf{Enj.}\\
        \hline
        P1 & -1 & 0 & 0 & 1 & -1 & -1 & 1 & -1 & 0 & -1 & -1 & -1 & 1 & 0 & -1 & 0 \\
        P2 & -1 & 0 & 1 & 1 & -1 & -1 & -1 & -1 & 1 & 1 & 1 & 1 & -1 & -1 & -1 & -1 \\
        P3 & -1 & 0 & 0 & -1 & 1 & 1 & 1 & 1 & 1 & 0 & 0 & 1 & -1 & -1 & -1 & -1 \\
        P4 & -1 & -1 & -1 & -1 & 1 & 1 & 1 & 1 & 0 & 1 & 0 & 0 & -1 & -1 & -1 & -1 \\
        P5 & -1 & -1 & -1 & -1 & 0 & 0 & 0 & -1 & 0 & 0 & 0 & -1 & -1 & -1 & -1 & -1 \\
        P6 & 1 & 1 & 1 & 1 & 0 & 1 & 1 & 1 & 1 & 1 & 1 & 1 & 0 & 1 & 0 & 1 \\
        \hline
        Avg. & -0.67 & -0.17 & 0.00 & 0.00 & 0.00 & 0.17 & 0.50 & 0.00 & 0.50 & 0.33 & 0.17 & 0.17 & -0.50 & -0.50 & -0.83 & -0.50 
    \end{tabular}
    \begin{tabular}{l|cccc|cccc|cccc|cccc}
        & \multicolumn{4}{c|}{\textbf{Temp. 1}} & \multicolumn{4}{c|}{\textbf{Com. 2}} & \multicolumn{4}{c|}{\textbf{Com. 0}} & \multicolumn{4}{c}{\textbf{I.S. 8}}\\
        & \textbf{Int.} & \textbf{Adpt.} & \textbf{Ctrl.} & \textbf{Enj.} & \textbf{Int.} & \textbf{Adpt.} & \textbf{Ctrl.} & \textbf{Enj.} & \textbf{Int.} & \textbf{Adpt.} & \textbf{Ctrl.} & \textbf{Enj.} & \textbf{Int.} & \textbf{Adpt.} & \textbf{Ctrl.} & \textbf{Enj.}\\
        \hline
        P1 & 1 & 0 & 1 & 1 & 0 & 1 & 1 & 1 & -1 & 0 & 1 & -1 & 1 & 1 & 1 & 1 \\
        P2 & 1 & -1 & 0 & -1 & 1 & 1 & 0 & 1 & 1 & 1 & -1 & 1 & -1 & -1 & -1 & -1 \\
        P3 & 0 & -1 & -1 & -1 & 1 & 0 & 0 & 1 & 1 & 1 & 1 & 1 & 0 & 0 & 1 & 0 \\
        P4 & -1 & -1 & -1 & -1 & 0 & -1 & -1 & -1 & 1 & 1 & 1 & 1 & 1 & 1 & 1 & 1 \\
        P5 & 1 & 0 & 1 & 1 & -1 & 1 & 0 & -1 & -1 & -1 & -1 & -1 & 1 & 1 & 1 & 1 \\
        P6 & -1 & -1 & 1 & 1 & 0 & 0 & -1 & -1 & 1 & 1 & -1 & -1 & -1 & -1 & -1 & 0 \\
        \hline
        Avg. & 0.17 & -0.67 & 0.17 & 0.00 & 0.17 & 0.33 & -0.17 & 0.00 & 0.33 & 0.50 & 0.00 & 0.00 & 0.17 & 0.17 & 0.33 & 0.33 
    \end{tabular}
    \caption{Users' intra-study experiment scores, where a user choosing the indicated setting gets a value of 1, choosing ``No Difference'' gets a value of 0, and choosing the alternative gets a value of -1.}
    \label{tab:quant_results_detailed}
\end{table*}

\end{document}